# Structural and Optical Studies on Manganese doped, Modified Sodium Potassium Lithium Niobate Lead –free Piezoelectric Ceramics

S. K. Mishra[1], R. K. Mishra[2], Kumar Brajesh[1], Rajyavardhan Ray[3], A. K. Himanshu[3], N. K. Singh[1]

[1]Department of Physics, Veer Kunwar Singh University, Ara, PIN- 802301, Bihar, India

[2] Department of Physics, Hindustan College of Science & Technology Farah, Mathura, 281122, U.P India

[3] Nanostructured & Advanced Materials Laboratory, Variable energy Cyclotron Centre, 1/AF, Bidhanngar, Kolkata-700064

**Abstract:**
Structural and Optical properties of the $(Na_{0.5}K_{0.5})_yLi_{1-y}NbO_3$ (y= 0.934 & 0.936) ceramics have been explored in terms of the X-Ray diffraction patters, Rietveld analysis, UV-Vis spectroscopy, and Kubelka-Munk functions. Rietveld analysis of both these structures reveals them to be tetragonal with space group P4mm. It is found that increase in Li doping increases the optical band gap. Presence of additional hardening dopant $MnO_2$ is found to have a uniform behavior for both these composition. With increasing $MnO_2$ concentration, the band gap decreases indicative of increase in B-site cation and O-*p* orbitals. Different valences of the Mn ions and relative difference in the ionic radii of the Mn ions and Nb ion are most likely the reason for these optical properties.

Keywords: Rietveld analysis, Band Gap, Kubelka-Munk

*Corresponding authors. Tel.: +91 33 23371230 (Ex:2408), Mobile.: +91 9883239177
Fax.: +91 33 23346871, Email : himanshu_ak@yahoo.co.in, akh@vecc.gov.in ( A. K. Himanshu).

## 1. Introduction:

Compounds with perovskite structure (ABO)$_3$ have attracted immense research interests because of the diversity of their physical properties exhibited by them. PbTiO$_3$-PbZrO$_3$ (PZT) based materials have been employed in many piezoelectric applications for the last six decades [1]. This is due to increasing needs of high-tech industries of new lead-free ferroelectric materials, which would reduce environmental contamination. ABO$_3$ type compounds have attracted attraction due to their relatively simpler structure. Sodium Niobate based solid solutions are interesting both from fundamental and practical view point. Pure NaNbO$_3$ has a perovskite structure and undergoes one of the most complicated sequences of phase transition as a function of temperature [2]. The Li doped Sodium Niobate system, with chemical formula (Li$_x$Na$_{1-x}$)NbO$_3$ (LNN) is of practical importance because of its low density (~ 4.5g/cm$^3$). LNN also have high sound speed ( > 6 Km/s), and wide range of curie temperatures, dielectric constants, and acceptable piezoelectric characteristic. Solid solution based on Potassium Sodium Niobate ( (K$_x$Na$_{1-x}$)NbO$_3$ ) are promising lead-free candidate for replacing the lead based piezoelectric ceramics. It is believed that in these ceramics exhibit remarkable properties close to the Morphotropic Phase Boundary (MPB).

Alkaline Niobate systems are, nowadays, considered as very promising candidates for technological applications. Compared to (Bi$_{0.5}$Na$_{0.5}$)TiO$_3$ (BNT) based system, the Alkaline Niobate systems are preferable as the former have high leakage current. It should be noted that the MPB is very different in the lead-based compounds in KNN-based ceramics because it shows strong temperature dependence [3], making them interesting candidate not only from the technological point view, but also as fundamental to the understanding of the inpact of doping (both A and B site), grain size, etc. on or around the MPB. .

From the technological point of view, in order to enhance different properties and sintering of these samples, additional hardening dopant is also used. Li, Ta, Zr, Ca, Mn, W based dopants have been used to obtain the optimal piezoelectric properties in these and PZT-based and other ceramics [13]. MnO$_2$ doping in (Na$_{0.5}$K$_{0.5}$)NbO$_3$ (NKN) has led to enhanced piezoelectric and electromechanical properties. These results are appreciable and comparable to the PZT ceramics [4-7].

It has earlier been reported that Alkaline Niobate systems, Li-doped NKN ceramics show relatively high curie temperatures ($T_c$) and good electrical properties [8]. A composition of $(Na_{0.5}K_{0.5})_{0.935}Li_{0.065}NbO_3$ (NKLN935) with a $T_c$ of 463 °C has been explored and is believed to lie close to MPB. The above authors have further studied the $MnO_2$ doped in NKLN935 and have reported the corresponding X-ray diffraction (XRD) patterns with peak-indexing as well as various electrical properties [9]. It was found that $MnO_2$ could be a useful dopant in NKLN systems for power applications, as it leads to harder ceramics with lower relative permeability and loss values, higher quality factors and electromechanical coupling factors. Therefore, in this article we concentrate on two neighboring compositions, $(Na_{0.5}K_{0.5})_yLi_{1-y}NbO_3$ (y= 0.934 & 0.936). We will explore the structural and optical properties of these compositions at y= 0.934 and 0.936, *i.e*, $(Na_{0.5}K_{0.5})_{0.934}Li_{0.066}NbO_3$ (NKLN934) and $(Na_{0.5}K_{0.5})_{0.934}Li_{0.066}NbO_3$ (NKLN936). In particular, we have carried out a detailed analysis of the structural properties through Reitveld analysis of X-Ray Diffraction (XRD) patterns.

For studying the optical properties of these compositions, measurements of diffuse reflectance of a powdered sample with a UV-Visible (UV-Vis) spectrophotometer is a standard technique. It is used for the determination of the absorption properties and optical band gap [10, 11]. We have performed the UV-Vis spectroscopy and experimentally determined the optical band gaps for these compositions along with the effect of $MnO_2$ doping on the band gap using the Kubelka-Munk Function.

2. **Experimental Details**

The polycrystalline samples of $(Na_{0.5}K_{0.5})_yLi_{1-y}NbO_3$ (y= 0.934 & 0.936) were prepared by solid-state reaction method. The starting materials were high purity (99.9%) powders of $KCO_3$, $KCO_3$ (99%), $Nb_2O_5$(99.9%) and $LiCO_3$ (99.99%), $MnO_2$ (99.99%). All these materials were dried at 150°C, dried in air, and then calcined at 900°C for 5 hours. After the calcination, the calcined powder was crushed. The crushed fine powders were mixed with a few drops of 2% polyvinyl alcohol as a binder in a mortar and pestle. The powder mixed with 2% PVA solution was cold- compacted in a steel die of 8mm diameter to give out circular disk shaped green pellets using a hydraulic pressure. The optimum compaction load was taken as 4 tons. Before firing, the green pellets were first annealed at $500^0$ C for 10 hours to burn of the binder (PVA). The sintering temperature, time and atmosphere play important roles in deciding the final property of the sintered ceramic body. The calcined powder was then mixed with a small amount of manganese oxide ($MnO_2$), represented by the formula (1-x) $(Na_{0.5}K_{0.5})_yLi_{1-y}NbO_3$ (y= 0.934 & 0.936) + x $MnO_2$ (x = 0.01, 0.015, 0.02). The $MnO_2$ added to $(Na_{0.5}K_{0.5})_yLi_{1-y}NbO_3$ (y= 0.934 & 0.936) were mixed for another10 hours to homogeneously distribute the dopant. The dried powder was compacted into disks in a die plunger having tungsten carbide tips. We put it in the furnace and these pellets were sintered at $1050^0$ C for $(Na_{0.5}K_{0.5})_yLi_{1-y}NbO_3$ (y= 0.934 & 0.936). Sintering was carried in the temperature range of 1050 -1070 $^0$C for the $MnO_2$ doped in $(Na_{0.5}K_{0.5})_yLi_{1-y}NbO_3$ (y= 0.934 & 0.936) in the programmable high temperature oven.

The sintered pellets were crushed to powder for the identification of the phase purity and crystal structure details of the samples using the X-ray diffraction (XRD; D-8 Advance Brukers).

Diffuse reflectance measurements of the powder samples were done with Perkin Elmer Lambda 950 UV-Vis-NIR spectrophotometer equipped with a 150 mm integrating sphere. The spectra were measured in the wavelength range 200-2500 nm. Spectralon was used as reference material.

### 3. Rietveld refinement details

Rietveld refinement of these compositions were carried out using the XRD data with a Rietveld refinement program FULLPROF [12]. The background was fitted with 6-Cofficient polynomial function, while the peak shapes were described by pseudo-Voigt profiles. In all the refinements, scale factor, lattice parameters, positional coordinates (*x,y,z*) and thermal parameters were varied. Occupancy parameters of all the ions were kept fixed during refinement. No correlation between the positional and thermal parameters was observed during refinement and as such it was possible to refine all the parameters together.

## 4. Results and discussion

The powder of $(Na_{0.5}K_{0.5})_yLi_{1-y}NbO_3$ (y= 0.934 & 0.936) compositions were prepared by calcining the stoichiometric amounts of various constituents at $900^0C$. These powders were sintered at 1050 -1070 $^0C$ to obtain the ceramic pellets of good quality and high densities. All the reflection peaks of the X-ray profile were indexed. Figs. 1 & 2 show the X-ray diffraction (XRD) patterns of the present compositions and all the reflection peaks of the X-ray profile are indexed. The structure is tetragonal (space group P4mm) for both the compositions NKLN934 & NKLN936. The $K^{1+}/Na^{1+}$ ions occupy 1 (a) sites at (0,0,z), $L^{1+}/Nb^{5+}$ and $O_I^{2-}$ occupy 1 (b) sites at (1/2,1/2,z), and $O_{II}^{2-}$ occupy 2(c) sites at (1/2,0,z). For the refinement, the initial values of the lattice parameters were obtained from our XRD data by least squares method. In this structure, $K^{1+}/Na^{1+}$ coordinates were fixed at (0, 0, 0) in our refinement. Fig. [1 & 2] depicts the observed, calculated and difference profiles for the refined structure after removing the peaks corresponding to the Cu-k$\alpha_2$ wavelength. The fit is found to be quite good. The refined structural parameters and the positional coordinates of these compositions for NKLN934 & NKLN936 are given in Table 1 & 2.

To analyze the structure of these compositions $(Na_{0.5}K_{0.5})_yLi_{1-y}NbO_3$ (y= 0.934 & 0.936), we have concentrated on the 200, 220, 222 pseudo-cubic X- ray diffraction profiles. It is known that, around the MPB composition, the structure is either rhombohedral, or tetragonal, or there is a coexistence of these two phases. In our case, we expect that the structure of $(Na_{0.5}K_{0.5})_yLi_{1-y}NbO_3$ (y= 0.934 & 0.936) compositions may be one of the above. For the rhombohedral structure, the 200 reflection appears as a singlet where as 220 and 222 appears as doublet with weaker reflection on the lower $2\theta$ side. For the rhombohedral structure, the 200 reflection peak appears as a singlet, where as the 220 and the 222 peaks appear as doublet with weaker reflection on the lower $2\theta$ side. For a tetragonal structure, while the 200-peak is a doublet with weaker reflection on the lower $2\theta$ side, the 220 peak is also a doublet with weaker reflection on the higher $2\theta$ side whereas the 222 peak is a singlet. For a monoclinic structure, all the three reflection peaks *viz.* the 200, 220, and 222 peaks show splitting [13]. For the present composition, we analyze this set of peak profiles. We find that the 200 peak is a doublet with weaker reflection on the lower 2θ side, the 220 peak is also a doublet with weaker reflection on the higher 2θ side, and the 222 peak is a singlet. Hence, it is evident that the structure of this composition is tetragonal, the space group being P4mm, as obtained from the Rietveld refinement.

Furthermore, we also find that the NKLN936 has more tetragonal distortion that the NKLN934 composition, as indicated by the corresponding *c/a* ratio. In the tetragonal distortion, the unit cell gets elongated along one direction, which leads to change in hybridization between the participating B-site cation and O atoms. This further leads to change in electronic and optical properties of the compound. The values of the *c/a* ratio for these compositions are found to be 1.028 and 1.035 for the NKLN934 and NKLN936 compositions, respectively. These changes are clearly exhibited in the optical band gap of these samples.

The Optical band gaps of $(Na_{0.5}K_{0.5})_yLi_{1-y}NbO_3$ (y= 0.934 & 0.936) and also the $MnO_2$ doped (at x = 0.01, 0.015, 0.02) samples were determined by the diffuse reflectance measurements. The, thus obtained, reflectivity spectra can be converted to an equivalent absorption spectra using the Kubelka–Munk (KM) function [14]. The KM function at any wavelength is given as:

$$F(R_\infty) = \frac{(1-R_\infty)^2}{2R_\infty} = \frac{\alpha}{s} \quad \ldots\ldots\ldots\ldots\ldots\ldots(1)$$

Here, $F(R_\infty)$ is the remission or Kubelka–Munk function, $R_\infty$ is the reflectance of the sample relative to the reference material $((R_\infty = R_{sample}/R_{splectron})$, α is the absorption coefficient, and s in the scattering coefficient. Typically, the scattering coefficient is only weakly dependent on energy. Therefore, $F(R_\infty)$ can be assumed to be proportional to the absorption spectrum [15]. The energy dependence of the adsorption coefficient near the absorption edge cane be expressed as

$$\alpha \propto \frac{(h\nu - E_g)^n}{h\nu}, \qquad \ldots\ldots\ldots\ldots(2)$$

as suggested by Davis and Mott [16]. Here $h\nu$ is the photon energy and $E_g$ is the energy of the optical transition corresponding to the optical band gap. Value of the exponent *n* depends on the nature of the optical transition and takes the value ½, 2 , 3/2, or 3 for direct-allowed, indirect-allowed, direct- forbidden or indirect forbidden transitions, respectively. With an appropriate value of *n*, the plot of *(αhυ)$^{1/n}$* vs hυ is linear near the absorption edge and the value of the optical band gap can be determined from the slope of the linear part. For the direct band gap $E_g$, we obtain a the expression

$$[F(R_\infty)h\nu]^2 = C(h\nu - E_g). \qquad \ldots\ldots\ldots\ldots(3)$$

Therefore, obtaining the $F(R_\infty)$ from Eq.1 and plotting the $[F(R_\infty)h\nu]^2$ against $h\nu$, the band gap $E_g$ of a powder sample can be extracted easily. Alternatively, it is also possible to obtain the optical band gap by directly plotting the appropriate power of the Kubelka-Munk, $[F(R_\infty)]^n$ as a function of the energy/frequency of incident photons, where *n* taked the values as above [17, 18]. We have used the second method to obtain the band gap for above compositions. The optical band gap values for the samples $(Na_{0.5}K_{0.5})_yLi_{1-y}NbO_3$ (y= 0.934 & 0.936)  and $MnO_2$ doped (at x = 0.01, 0.015, 0.02) for n=2 is given in the Table 3 & 4. Figs. 3 and 4, respectively, show the behavior of $[F(R_\infty)h\nu]^2$ with increasing energy of the incident photon, $h\nu$. We have plotted the corresponding functions for all the above values of *n* (not shown). However, the best fit was found to be for the value of *n=2*. This implies that both the above compositions hsave direct band gaps. Due to insulating behavior of these compositions, the values of the plotted function is zero in the low energy limit and has linearly increasing non-zero values for energy values greater than a threshold energy of the incident photon.  As mentioned above, the value of the optical band gap is obtained from intercept of low energy linear behavior of this function, as also shown in the corresponding figures.

A comparison of the intercept for pure NKLN934 (Fig. 3a) and NKLN936 (Fig. 4a) show that with increasing Li doping, the band gap increases. The corresponding values of the band gap are 3.22eV and 3.32eV, respectively. This is primarily due to increase in tetragonality of these compositions with increase in Li doping, as mentioned above. This increase in tetragonality leads to larger overlap and, therefore, stronger hybridization of Nb-*d* and O-*p* orbitals. The distribution of the O-*p* orbitals is not isotropic around the Nb atom. There is stronger hybridization between the Nb-dz2 orbital and the pz orbital of the O1 atoms located at (0,0,*z*). Similarly, there is also stronger hybridization between the planar *d* orbitals and the planar O-*p* orbitals, *px* and *py*.

Interestingly, with MnO2 doping, the band gaps for both NKLN934 and NKLN936 compositions show uniform trend. Tables 3 & 4, respectively, show the values of the band gap of these compositions, as obtained from the intercepts in Figs. 3 and 4 (Panels (b)-(d)). It is interesting to note that the band gap decreases with increasing concentrations of MnO2. It is usually considered that the Mn ions occupy the B-site. However, it is not clear whether this substitution for Nb5+ would lead to A-site cation vacancies or Oxygen vacancies to be produced. This situation if further compounded by multiple valence states and large difference in the ionic radii of the Mn and Nb ions. Mn ions are present in both +2 and +4 state, which have the ionic radii of 0.67nm and 0.54 nm. The ionic radius of the Nb5+ ion, on the other hand, is 064nm. This difference in valency and ionic radii will leads to distortion of the O octahedra surrounding the Nb and Mn ions, eventually leading to different hybridization between the O and B-site ions. For the NKLN935 composition, it has been argued that MnO2 doping leads to formation of osygen vacancies. As can be observed from the band gap values, the effective tetragonality of the pure NKLN934 and KNLN936 composition seems to decrease with increase in MnO$_2$ doping, and is likely to be due to increase in hybridization between the B-site ion and the *pz* orbital of the O1 atoms.

This increase in band gap values is also indicative of enhanced piezoelectric properties of these compositions. Therefore, it is expected that, similar to the NKLN935 [9], MnO$_2$ doping may lead to decrease in dielectric constant, loss tangent and piezoelectric constant d$_{33}$, and increase in the mechanical quality factor Q$_m$. Therefore, it is certainly of interest to explore these compositions further, in particular, the dielectric and piezoelectric parameters. Furthermore, other similar compositions should also be explored to locate the

MPB as they seem to be excellent and very promising materials for piezoelectric transformer applications [13].

**Conclusions**

In conclusion, we have studied the structural and optical properties of the $(Na_{0.5}K_{0.5})_yLi_{1-y}NbO_3$ (y= 0.934 & 0.936) ceramics (respectively called NKLN934 and NKLN936) and found that both these compositions are tetragonal, similar to the previously studied $(Na_{0.5}K_{0.5})_{0.935}Li_{0.065}NbO_3$ (NKLN935). The effect of tetragonality, as obtained from the *c/a* ratio, shows that tetragonality increases with Li doping. This further leads to an increase in optical band gap values in these values. Detailed exploration of the optical properties, in terms of the UV-Vis spectrum and Kubelka-Munk function, reveal that the values of the optical band gap is, respectively, 3.22 eV and 3.32 eV, as expected from the relative change in hybridization between the B-site cation and O-p orbital. Furthermore, the $MnO_2$ doping shows a uniform trend for both the systems: the band gap decreases with increase in $MnO_2$ doping. This is clearly indicative of the promise of enhanced piezoelectric properties of theses compositions, as required for technological applications like piezoelectric transformers.

**Acknowledgements:**

One of the author **AKH is** thankful to Prof. D. K. Srivastava, Director of VECC, for his keen interest and encouragement in the XI-year plan conducting polymer project (PIC No. 11-R&D-VEC-5.09.2000.

**Table caption:**

**Table 1.**
Refined structural parameters of NKLM 934 using tetragonal (space group; P4mm) model.

| Ions | Positional coordinates | | | thermal parameters |
|---|---|---|---|---|
| | x | y | z | B(Å$^2$) |
| Na$^{1+}$/K$^{1+}$ | 0.00 | 0.00 | 0.00 | 0.0717(5) |
| Li$^{1+}$/Nb$^{5+}$/ | 0.50 | 0.50 | 0.5453(5) | 0.0382(3) |
| O$_I^{2-}$ | 0.00 | 0.50 | 0.477(5) | 0.0418(1) |
| O$_{II}^{2-}$ | 0.50 | 0.50 | 0.087(2) | 0.0022(1) |

a = b = 3.9622(4)Å, c = 4.074(5) Å, R$_P$ = 6.63, R$_{wp}$ = 8.89, R$_{exp}$ = 5.43, R$_B$ = 7.90, R$_f$ = 8.94, Vol. = 62.911(1) and $\chi^2$ = 2.68

**Table 2.**
Refined structural parameters of NKLM 936 using tetragonal (space group; P4mm) model.

| Ions | Positional coordinates | | | thermal parameters |
|---|---|---|---|---|
| | x | y | z | B(Å$^2$) |
| Na$^{1+}$/K$^{1+}$ | 0.00 | 0.00 | 0.00 | 0.31(2) |
| Li$^{1+}$/Nb$^{5+}$/ | 0.50 | 0.50 | 0.51897(1) | 0.042(3) |
| O$_I^{2-}$ | 0.00 | 0.50 | 0.46(1) | 0.7595(1) |
| O$_{II}^{2-}$ | 0.50 | 0.50 | 0.08733(1) | 0.145(5) |

a = b = 3.961196(2)Å, c = 4.09995(2) Å, R$_P$ = 11.7, R$_{wp}$ = 14.7, R$_{exp}$ = 5.95, R$_B$ = 5.43, R$_f$ = 3.24, Vol. = 62.921(1) and $\chi^2$ = 6.09

Table 3.

| Sample Name | Direct Band gap (eV) |
|---|---|
| NKLN 934  [Fig. 3(a)] | 3.22 |
| NKLN 934-MnO$_2$ 0.01  [Fig. 3(b)] | 3.21 |
| NKLN 934-MnO$_2$0.015 [Fig. 3 (c)] | 3.20 |
| NKLN 934-MnO$_2$0.02  [Fig. 3(d)] | 3.18 |

**Table 4:**

| Sample Name | Direct Band gap (eV) |
|---|---|
| NKLN 936 [Fig. 5(a)] | 3.32 |
| NKLN 936-$MnO_2$ 0.01 [Fig. 5(b)] | 3.30 |
| NKLN 936-$MnO_2$ 0.015 [Fig. 5(c)] | 3.28 |
| NKLN 936-$MnO_2$ 0.02 [Fig. 5(d)] | 3.26 |

**Figure Caption:**

1. Observed (circle), Calculated (continuous lines), and difference (bottom of the figure) profiles in the 2θ range 20-80 degree for the given composition for the NKLN934.
2. Observed (circle), Calculated (continuous lines), and difference (bottom of the figure) profiles in the 2θ range 20-80 degree for the given composition for the NKLN936.
3. Kubelka –Munk transformed reflectance spectra of function; (a) NKLN934, (b) NKLN 934-$MnO_2$ 0.01 , (c) NKLN 934-$MnO_2$ 0.015, (d) NKLN NKLN 934-$MnO_2$ 0.02
4. Kubelka –Munk transformed reflectance spectra of function; (a) NKLN936, (b) NKLN 936-$MnO_2$ 0.01 , (c) NKLN 936-$MnO_2$ 0.015, (d) NKLN NKLN 936-$MnO_2$ 0.02

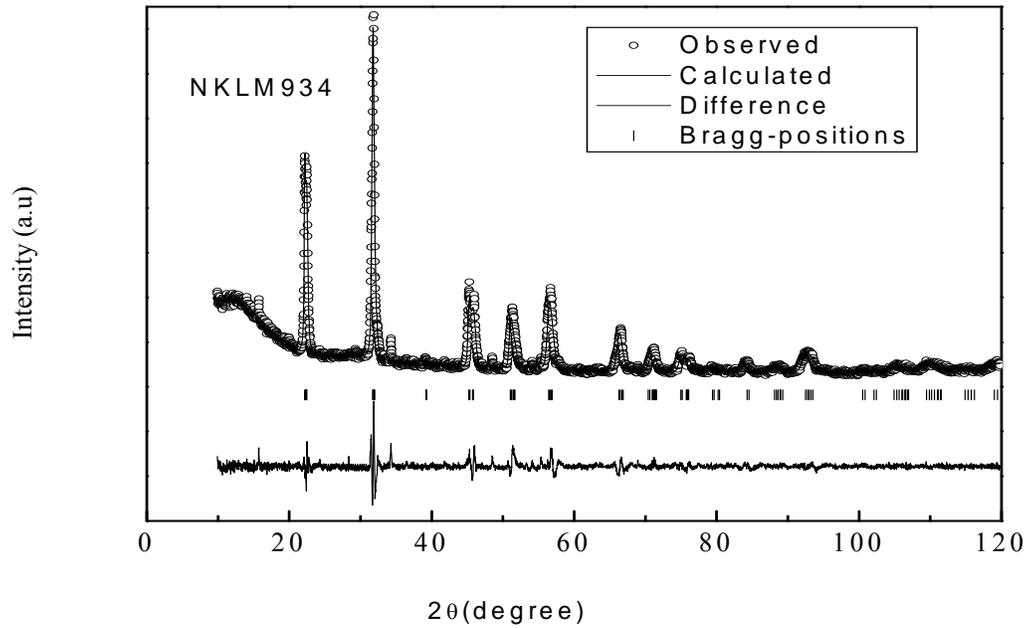

**Fig.1** Observed (circle), Calculated (continuous lines), and difference (bottom of the figure) profiles in the 2θ range 20-80 degree for the given composition for the NKLN934.

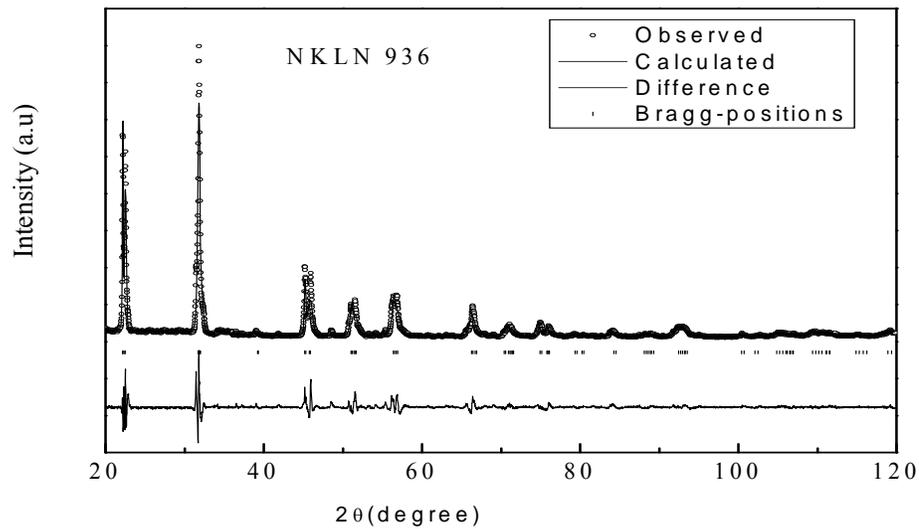

Fig.2: Observed (circle), Calculated (continuous lines), and difference (bottom of the figure) profiles in the 2θ range 20-80 degree for the given composition for the NKLN936.

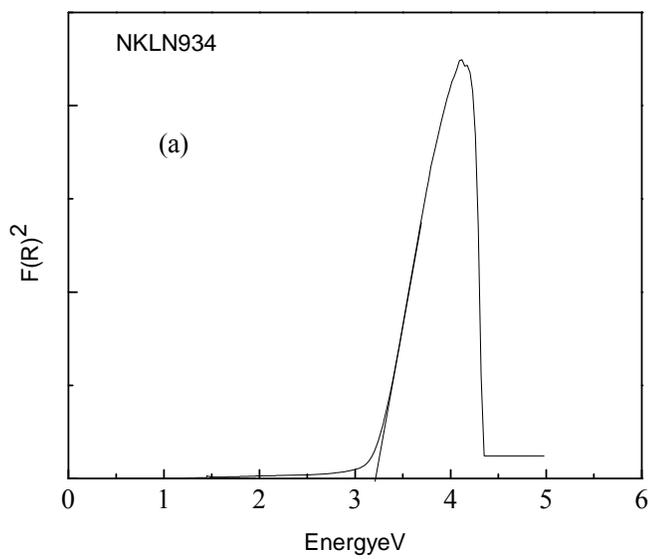

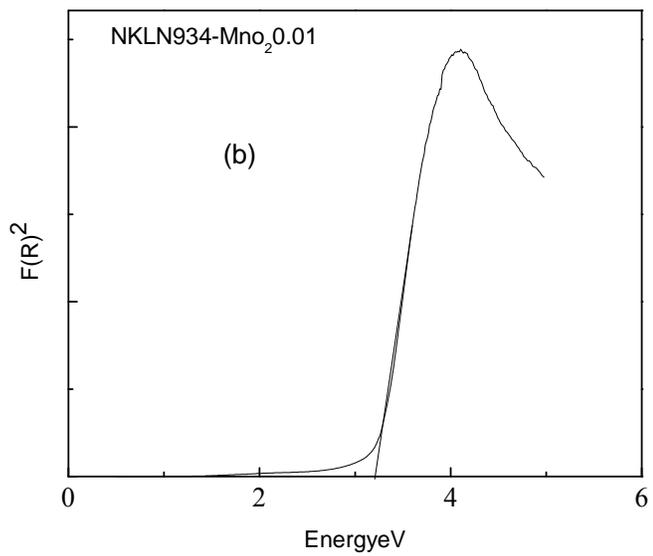

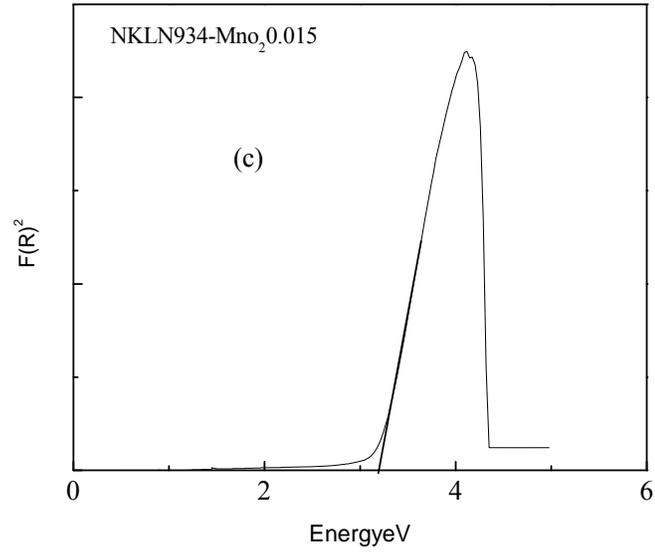

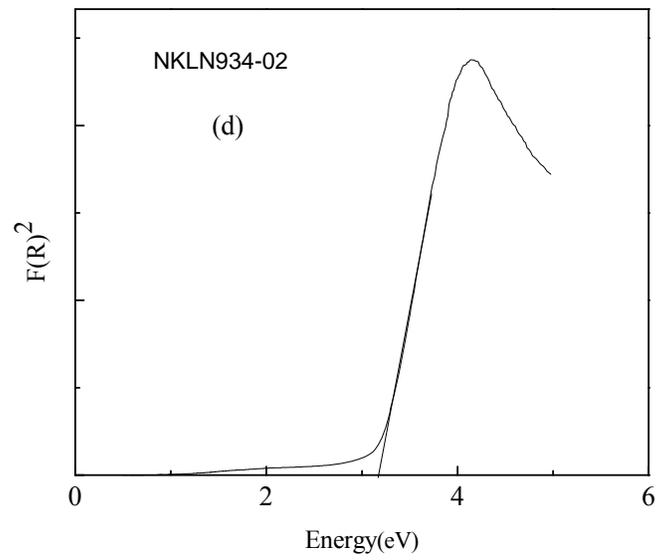

**Fig.3:** Kubelka –Munk transformed reflectance spectra of function; (a) NKLN934, (b) NKLN 934-MnO$_2$ 0.01 , (c) NKLN 934-MnO$_2$ 0.015, (d) NKLN NKLN 934-MnO$_2$ 0.02

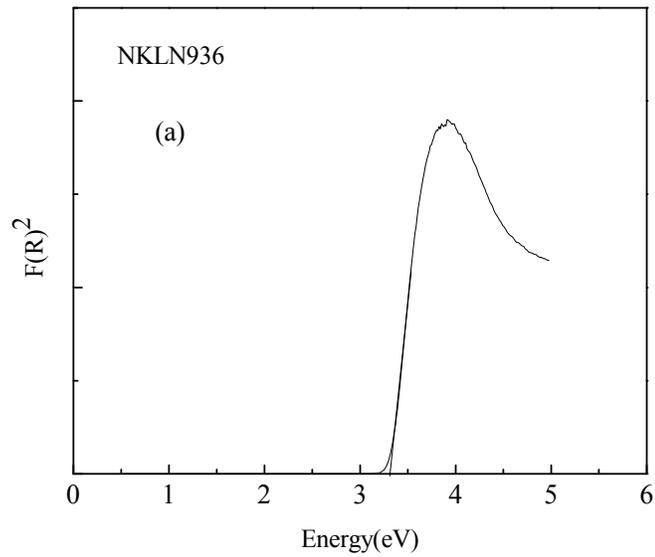

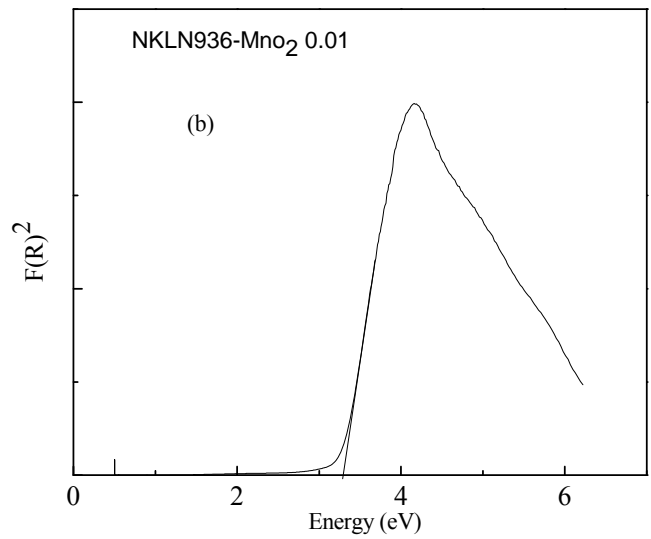

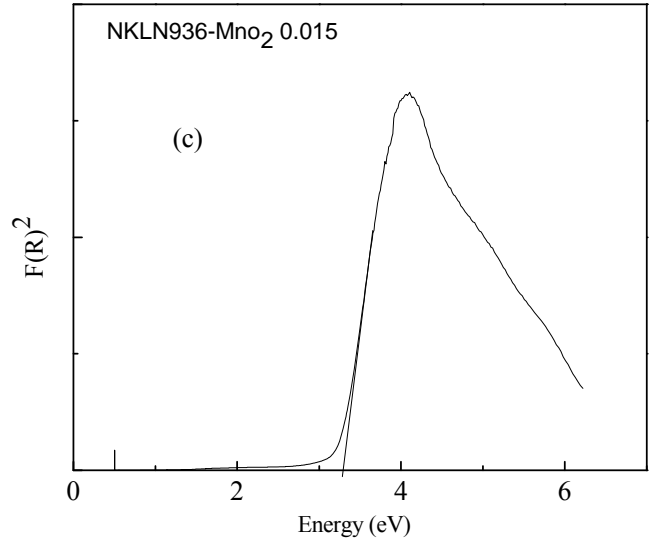

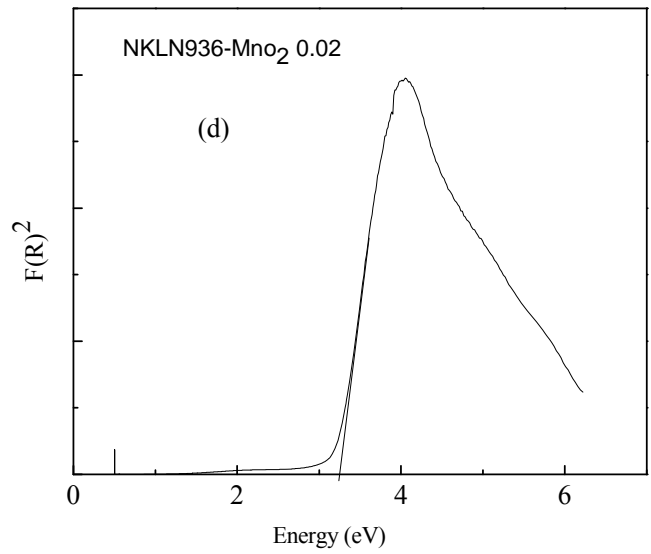

**Fig.4:** Kubelka –Munk transformed reflectance spectra of function; (a) NKLN934, (b) NKLN 934-MnO$_2$ 0.01 , (c)  NKLN 934-MnO$_2$ 0.015, (d) NKLN NKLN 934-MnO$_2$ 0.02